\documentclass[preview,11pt]{article}

\usepackage{amssymb,latexsym}
\usepackage{graphicx}
\usepackage{amsfonts, amsmath, amssymb, dsfont, setspace}
\usepackage{fixltx2e}

\usepackage{color}
\usepackage{verbatim}
\usepackage{vmargin}
\usepackage{url}
\usepackage[english]{babel}
\usepackage[latin1]{inputenc}
\usepackage[T1]{fontenc}
\usepackage{enumitem}
\usepackage{array}
\usepackage{fancyhdr}
\usepackage{lmodern}
\usepackage{hyperref}
\usepackage{cleveref}
\crefformat{equation}{(#2#1#3)}
\usepackage[font={footnotesize}]{caption}

\begin{document}

\thispagestyle{empty}

\title{Spatiotemporal stability of periodic travelling waves in a heteroclinic-cycle model}
\date{}
      \maketitle



\begin{center}
\vspace{2mm}
Cris R. Hasan\footnote{School of Mathematical Sciences, University College Cork, Ireland; Corresponding author: \url{rhasan@ucc.ie}.}, Hinke M. Osinga\footnote{Department of Mathematics, University of Auckland, New Zealand.}, Claire M. Postlethwaite\footnotemark[2], Alastair M. Rucklidge\footnote{Department of Mathematics, University of Leeds, UK.}

\end{center}

\subsubsection*{Abstract}
\begin{small}
 We study a Rock--Paper--Scissors model for competing populations that exhibits travelling waves in one spatial dimension and spiral waves in two spatial dimensions. A characteristic feature of the model is the presence of a robust heteroclinic cycle that involves three saddle equilibria. The model also has travelling fronts that are heteroclinic connections between two equilibria in a moving frame of reference, but these fronts are \emph{unstable}. However, we find that large-wavelength travelling waves can be \emph{stable} in spite of being made up of three of these unstable travelling fronts. In this paper, we focus on determining the essential spectrum (and hence, stability) of large-wavelength travelling waves in a cyclic competition model with one spatial dimension. We compute the curve of transitions from stability to instability with the continuation scheme developed by Rademacher et al. (2007 \emph{Physica~D} 
{\bf 229} 166--83). We build on this scheme and develop a method for computing what we call \emph{belts of instability}, which are indicators of the growth rate of unstable travelling waves. Our results from the stability analysis are verified by direct simulation for travelling waves as well as associated spiral waves. We also show how the computed growth rates accurately quantify the instabilities of the travelling waves.
\end{small}

\vspace{2pc}
\noindent{\it Keywords}: stability of travelling waves, spiral waves, heteroclinic cycles, rock--paper--scissors

\pagebreak
\section{Introduction}

 Cyclic interactions of three competing populations have been observed in various ecosystems including morphs of the side-blotched lizard \cite{Sinervo1996, Sinervo2000}, coral reef invertebrates~\cite{Jackson1975} and strains of \emph{Escherichia coli}
 \cite{Kerr2002, Kirkup2004}.
The famous game of Rock-Paper-Scissors, where Rock crushes Scissors, Scissors cut Paper and Paper wraps Rock, provides a model for cyclic dominance between competing populations in ecology, or strategies in evolutionary game theory.
This phenomenon can be described by the May--Leonard model~\cite{May1975}, a system of ordinary differential equations (ODEs) that are defined as follows:

%
\begin{equation}
\label{eq:MayLeonard}
  \left\{
    \begin{array}{rrlll}
      \dot{a}& = & a \, (1-a-b-c-(\sigma+\zeta)b+\zeta c),    \\
      \dot{b}& = & b \, (1-a-b-c-(\sigma+\zeta)c+\zeta a),    \\
      \dot{c}& = & c \, (1-a-b-c-(\sigma+\zeta)a+\zeta b).
    \end{array}
  \right.
\end{equation}
 Here, the dot represents the derivative with respect to time $t$, the variables $a$, $b$ and $c$ are the (non-negative) densities of the three competing populations, non-dimensionalized to vary between 0 and 1, and $\sigma$ and $\zeta$ are parameters that describe how the species interact, assuming symmetry between the three species. This system has three on-axis equilibria $(a,b,c) = (1,0,0)$, $(0,1,0)$ and $(0,0,1)$, which are connected in a closed circuit via heteroclinic connections to form a so-called heteroclinic cycle. Since the heteroclinic connections between the pairwise equilibria lie in invariant planes, namely, $\{c = 0\}$, $\{a = 0\}$ and $\{b = 0\}$, respectively, this heteroclinic cycle is robust, that is, it persists under small perturbations that preserve the invariant subspaces~\cite{postlethwaite2017}.

In this paper, we assume that the competing populations are spatially distributed on the line or plane. This implies that the ODEs~\cref{eq:MayLeonard} become a system of partial differential equations (PDEs), such as the model presented in~\cite{Frey2010, reichenbach2007}:
\begin{equation}
  \label{eq:MeanField}
  \left\{
    \begin{array}{rrlll}
      \dot{a}& = &  a \, (1-a-b-c-(\sigma+\zeta)b+\zeta c)+ \nabla^2 a,	  \\
      \dot{b}& = &  b \, (1-a-b-c-(\sigma+\zeta)c+\zeta a)+ \nabla^2 b,	  \\
      \dot{c}& = &  c \, (1-a-b-c-(\sigma+\zeta)a+\zeta b)+ \nabla^2 c,
    \end{array} \right.
\end{equation}
where $a(x,y,t)$, $b(x,y,t)$ and $c(x,y,t)$ now depend on spatial coordinates as well as time. As in~\cref{eq:MayLeonard}, the dot represents the derivative with respect to time~$t$ and the Laplace operator~$\nabla^2$ ($= \frac{\partial^2}{\partial x^2} + \frac{\partial^2}{\partial y^2}$ in two dimensions) models diffusion of the species. We remark that the system of PDEs~\cref{eq:MeanField} can be derived from a stochastic model of interacting populations in an appropriate limit~\cite{Frey2010, Szczesny2013, Szczesny2014}.

A natural question to ask is: what happens to the heteroclinic cycle in the ODEs once spatial structure is added? In addition to spatially uniform and heteroclinic cycle solutions, the system of PDEs~\cref{eq:MeanField} is known to feature one-dimensional, periodic travelling wave (TW) solutions as well as (in two dimensions) spiral wave solutions and chaotic or disordered spatiotemporal patterns~\cite{reichenbach2007, Szolnoki2014}. 

\begin{figure}[t] 
  \centering
  \includegraphics{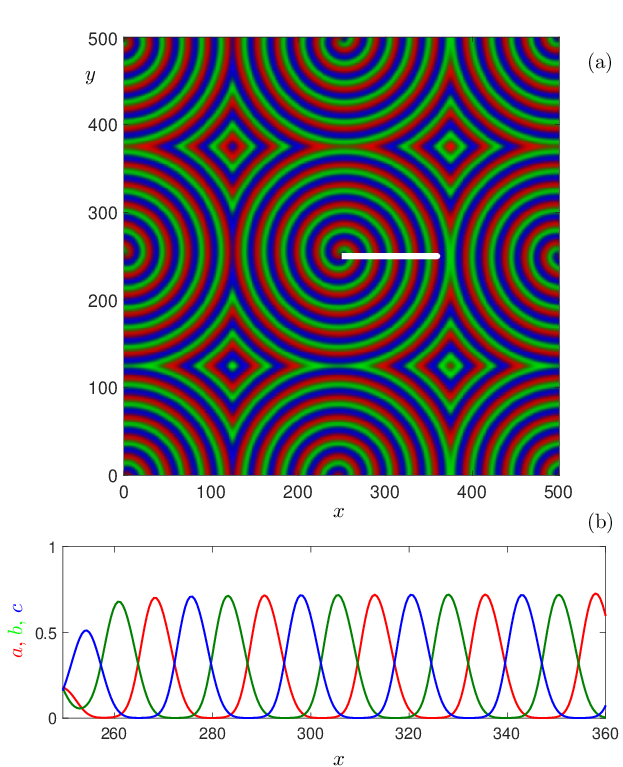}
  \caption{\label{fig:SWs}
    Spiral waves and their relation to travelling waves in system~\cref{eq:MeanField}, with $\sigma=3.2$ and $\zeta=0.8$, in a $500 \times 500$ domain with periodic boundary conditions. Here, red, green and blue represent $a$, $b$ and $c$, respectively. (a)~The central spiral wave rotates clockwise, in the $(x,y)$-plane. (b)~The values of $a$, $b$ and $c$ along the white line segment in panel~(a), with the core of the spiral on the left and (effectively) one-dimensional travelling waves on the right.
}
\end{figure}
%
\Cref{fig:SWs} shows a snapshot of spiral waves of system~\cref{eq:MeanField}, with $\sigma=3.2$  and $\zeta=0.8$ in a two-dimensional domain with periodic boundary conditions. The three populations chase each other in a rotational way in the clockwise direction: $a$~(red) chases $c$ (blue), $c$~chases $b$ (green) and $b$ chases~$a$. \Cref{fig:SWs}(b) illustrates the distribution of $a$, $b$ and $c$ along the white horizontal  cross-section shown in panel~(a). The solution along the cross-section appears to approach a periodic TW asymptotically as it propagates away from the central core of the spiral. As the TWs move along the domain in the positive $x$-direction, we observe the same chasing dynamics between the species $a$, $b$ and~$c$.
Postlethwaite and Rucklidge~\cite{postlethwaite2017, postlethwaite2019} illustrated that one-dimensional solutions of system~\cref{eq:MeanField} on periodic domains also show stable TWs with a variety of spatial periodicities and wavelengths, as well as other spatiotemporal behaviour.

In this paper, we complement the work done in~\cite{postlethwaite2017, postlethwaite2019} and focus on stability analysis of large-wavelength periodic TWs near a heteroclinic cycle.
In large one-dimensional periodic domains, the stable TWs take the form of three fronts connecting three equilibria of system~\cref{eq:MayLeonard}.
In an infinite domain, a single travelling front, connecting one equilibrium to another, is necessarily \emph{unstable}, since one of the connecting equilibria is unstable~\cite{sandstede2002stability}. 
We can understand this heuristically as follows. Restrict the dynamics to the invariant subspace with $c=0$, and consider the front between the equilibrium with $b=1$ and the equilibrium with $a=1$, in an  unbounded one-dimensional domain. Then, $a\rightarrow 0$ and $b\rightarrow 1$ as $x\rightarrow -\infty$, and $a\rightarrow 1$ and $b\rightarrow 0$ as $x\rightarrow +\infty$. The front moves from left to right: $b$ is out-competing $a$. Now, consider an arbitrarily small perturbation in $b$. In a background with $a=1$, such a perturbation will grow, and since the front moves at a constant, finite speed, we can make this perturbation sufficiently far to the right so that it grows before the front reaches it. Thus, the front is unstable.
However, a similar situation for a periodic travelling wave solution on a possibly large, but finite domain will see the front catch up with, and possibly absorb, any small, growing perturbation.
Indeed, we observe the counter-intuitive phenomenon where TWs made up of three arbitrarily long \emph{unstable} fronts cycling between three equilibria can be \emph{stable}.


Previous studies have shown that pulses formed by gluing together two unstable fronts can be stable \cite{Nii2000,romeo2000,sandstede2000Gluing}---a phenomenon that was described at the time as unexpected~\cite{Nii2000,sandstede2000Gluing} and paradoxical~\cite{romeo2000}.
Stability analysis of large-wavelength periodic TWs near heteroclinic cycles that are composed of three consecutive unstable fronts, to our knowledge, has not been done before.

For the stability analysis, we adapt the continuation-based numerical method described in~\cite{rademacher2007computing} to compute the spectrum of periodic TWs and obtain the stability boundaries in the parameter plane. The spatially-extended system~\cref{eq:MeanField} poses particular challenges because of the large dynamic range of the three variables over the heteroclinic cycle and the large period of the~TWs. 
This paper provides a blueprint on how to perform these technically challenging computations.
We also provide a demo via the supplementary data link \url{https://github.com/CrisHasan/TW_Nonlinearity2021} for computing essential spectra and stability boundary in parameter space using the open-source continuation software package {\sc Auto}~\cite{Doedel}.

We compare our stability calculations to direct simulations of system~\cref{eq:MeanField} and use the stability analysis for one-dimensional TWs to inform numerically determined stability boundaries for the two-dimensional spiral waves.
To define and quantify spatiotemporal instabilities, we introduce heuristic criteria for the destablization of TWs and spiral waves in the PDE simulations.
We find the instabilities of periodic TWs and spiral waves to be accurately captured by the stability analysis.

In direct simulations of system \cref{eq:MeanField}, large-wavelength periodic TWs with unstable essential spectra persist and remain stable for a very long time, even after introducing small perturbations.
We observe that this surprising phenomenon occurs when the growth rates of these TWs are very small.
This motivated us to expand on the work of~\cite{rademacher2007computing} and develop a technique for the numerical continuation of the growth rate of unstable periodic travelling waves.
The computed growth rates accurately and efficiently predict the computationally expensive integration time needed for the destabilization of~TWs.

The outline of the paper is as follows. \Cref{sec:existence} reviews the dynamics of system~\cref{eq:MeanField} in a travelling frame of reference and discusses the existence of periodic~TWs. In \cref{sec:stability}, we describe and implement the numerical scheme for computing the essential spectrum of periodic TWs and determining the boundary of the stability region. We then show in \cref{sec:simulations} how the spatiotemporal behaviour of travelling waves in the PDE simulations agrees with the acquired results of the stability analysis. We also find that the computed growth rate of the periodic TWs obtained from the spectral analysis serves as an accurate predictor and indicator of the instabilities of the TW solutions. Conclusions and final remarks are presented in \cref{sec:discussion}.

\section{Existence of periodic travelling waves} \label{sec:existence}
We use a dynamical systems approach to study the existence and stability of TW solutions of system~\cref{eq:MeanField}. Our starting point is the analysis in~\cite{postlethwaite2017, postlethwaite2019} that establishes the existence of TWs for system~\cref{eq:MeanField} with one spatial dimension; this section briefly reviews the relevant results from~\cite{postlethwaite2017, postlethwaite2019}. Writen in vector form, system~\cref{eq:MeanField} is given by
\begin{equation}
  \label{eq:vectorForm}
  \mathbf{U}_t=\mathbf{f(U)}+\mathbf{U}_{xx},
\end{equation}
where $\mathbf{U}(x,t)=(a(x,t),b(x,t),c(x,t))^T$, for $t, x \in \mathbb{R}$. The subscripts denote the partial derivatives, and $\mathbf{f(U)}$ represents the kinetic terms of~\cref{eq:MeanField}. We assume that the travelling waves move with a constant wavespeed $\gamma>0$ and introduce the change of coordinates $z=x+\gamma t$, so that $\frac{\partial}{\partial x} \mapsto \frac{\partial}{\partial z}$ and $\frac{\partial}{\partial t} \mapsto \gamma \frac{\partial}{\partial z} + \frac{\partial}{\partial t}$. The system in the travelling frame then becomes
\begin{equation}
  \label{eq:Tframe}
  \mathbf{U}_t=-\gamma \mathbf{U}_z +\mathbf{f(U)}+\mathbf{U}_{zz}.
\end{equation}
Stationary solutions of~\cref{eq:Tframe} can be obtained by setting $\frac{\partial}{\partial t}=0$, resulting in a 
second-order ODE that can be written as a six-dimensional system of first-order ODEs for $\mathbf{U}$ and~$\mathbf{U}_z$:
\begin{equation}
  \label{eq:6ODEs}
  \left\{
    \begin{array}{lll}
      \mathbf{U}'&= & \mathbf{U}_z, \\
      \mathbf{U}_z' & = &   \gamma    \mathbf{U}_z - \mathbf{f}(\mathbf{U}),
    \end{array} \right.
\end{equation}
where the prime denotes derivation with respect to~$z$.

System~\cref{eq:6ODEs} admits five non-negative equilibria: the origin $(a,b,c,a_z,b_z,c_z)= (0,0,0,0,0,0)$, the coexistence equilibrium $\frac{1}{\sigma+3}(1,1,1,0,0,0)$, and the on-axis equilibria $\xi_1=~ (1,0,0,0,0,0)$, $\xi_2=(0,1,0,0,0,0)$ and $\xi_3=(0,0,1,0,0,0)$. For all $\gamma,\zeta,\sigma>0$, there exists a robust heteroclinic cycle between the saddle on-axis equilibria~\cite{postlethwaite2017, postlethwaite2019} as follows. Consider the four-dimensional invariant subspaces $P_1:=\{c=c_z=0\}$, $P_2:=\{a=a_z=0\}$, and $P_3:=\{b=b_z=0\}$. For the dynamics restricted to each invariant subspace $P_i$, $i \in \{1,2,3\}$, the saddle equilibrium $\xi_i$ has a three-dimensional unstable manifold and saddle equilibrium $\xi_{i+1}$ (where $\xi_4 \equiv \xi_1$) has a two-dimensional stable manifold; these manifolds generically intersect pairwise in the four-dimensional invariant subspace $P_i$. Hence, there exists a one-dimensional heteroclinic connection $\xi_i \to \xi_{i+1}$ for all $P_i$ even though $\xi_i$ and $\xi_{i+1}$ are both saddles with respect to the invariant subspace $P_i$. The concatenation of the three heteroclinic
connections (travelling fronts) forms a robust heteroclinic cycle in the sense that it persists under perturbations that respect the invariant subspaces.

\begin{figure}[t] 
  \centering
  \includegraphics{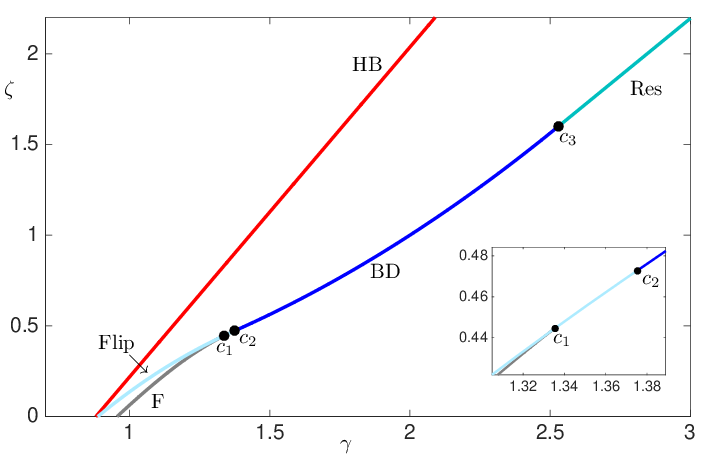}
  \caption{\label{fig:ODEbif}
    Bifurcation diagram of system~\cref{eq:6ODEs} in the $(\gamma,\zeta)$-parameter plane with $\sigma=~3.2$. The red and grey curves are the loci of Hopf bifurcation (HB) and fold bifurcation of periodic orbits (F), respectively. The differently shaded blue curves, from left to right, are the loci of heteroclinic flip bifurcation (Flip), Belyakov--Devaney-type heteroclinic bifurcation (BD) and resonance heteroclinic bifurcation (Res). The three dots labelled $c_1$, $c_2$ and $c_3$ are codimension-two points at which bifurcations of different types meet; the inset shows an enlargement near $c_1$ and $c_2$.}
\end{figure}
%
\Cref{fig:ODEbif} shows the bifurcation diagram of the travelling-frame system~\cref{eq:6ODEs} for varying $\gamma$ and $\zeta$, with $\sigma=3.2$. The red line is the locus of Hopf bifurcation~(HB) and the grey curve is the locus of fold bifurcation of periodic orbits~(F). The three blue-shaded curves are three different types of heteroclinic bifurcations: heteroclinic flip bifurcation~(Flip), Belyakov--Devaney-type heteroclinic bifurcation~(BD) and resonance heteroclinic bifurcation~(Res). The flip bifurcation in system~\cref{eq:6ODEs} is a degenerate heteroclinic bifurcation where each heteroclinic orbit is no longer tangent to the weak direction of the unstable manifold, but rather, tangent to the subspace spanned by the strong directions. We trace this bifurcation using a two-point boundary value problem set-up in conjunction with numerical continuation. The bifurcation~BD arises when two of the expanding eigenvalues of the saddle equilibria are equal, and the resonance heteroclinic bifurcation occurs when the heteroclinic orbit involves two eigenvalues of the same magnitude but opposite signs. The codimension-two bifurcation points $c_1, c_2$ and $c_3$ (black dots) are the meeting points of bifurcation curves~F and~Flip, of curves~BD and~Flip, and of curves~Res and~BD, respectively. Note that in \cite{postlethwaite2019}, preliminary numerical results suggested that $c_1=c_2$. Here, we have repeated the calculations more carefully and show that this is generally not the case.

A two-parameter family of periodic orbits born at the Hopf bifurcation exists to the right of the curve~HB. Another family of periodic orbits originates from the flip bifurcation and exists between the curves~Flip and~F. The two families collide and annihilate each other at the fold bifurcation of periodic orbits. Away from~F, the periodic orbits emanating from the curve HB terminate at one of the three heteroclinic bifurcation branches. There exists a single periodic orbit for each parameter value in the region in between HB and the heteroclinic bifurcations, and two periodic orbits co-exist in the region between~F and~Flip. Each periodic orbit with period $L$ in the travelling-frame system~\cref{eq:6ODEs} corresponds to a TW solution in system~\cref{eq:vectorForm} that is spatially periodic with wavelength $L$. 
We find that the codimension-two bifurcation points $c_3$ and $c_1$, related to degenerate heteroclinic bifurcations of system~\cref{eq:6ODEs}, also act as end points of a curve that marks the boundary of the parameter region for which the periodic TWs of system~\cref{eq:vectorForm} are stable; see already 
\cref{fig:EckBif1,fig:EckBif2,fig:EckBif3}.

\section{Stability analysis of periodic travelling waves}
\label{sec:stability}
 In this section, we perform a linear stability analysis of the periodic TWs. We first derive the eigenvalue problem and describe the computational set-up for computing spectra of TWs, which is adapted from~\cite{rademacher2007computing, sandstede2000absolute}. We then implement this analysis to determine the stability of periodic TWs in the spatially-extended Rock--Paper--Scissors model ~\cref{eq:MeanField}. The Python drivers for these {\sc Auto} calculations are available in the form of a demo via the supplementary data link (\url{https://github.com/CrisHasan/TW_Nonlinearity2021}).

\subsection{Derivation of the eigenvalue problem}
We assume that $\mathbf{\widehat{U}}$ is a periodic TW solution with spatial period or wavelength $L$. This TW is required to be a stationary solution to the PDE in the travelling frame given by system~\cref{eq:Tframe}, that is,
\begin{equation*} 
  -\gamma \mathbf{\widehat{U}}_z +\mathbf{f(\widehat{U})}+\mathbf{\widehat{U}}_{zz}=0.
\end{equation*}
Hence, in the travelling-frame coordinates, $\mathbf{\widehat{U}}$ depends only on $z$ and $\mathbf{\widehat{U}}(z + L) = \mathbf{\widehat{U}}(z)$ for all $z \in \mathbb{R}$. 
Now consider the perturbation $\mathbf{U}(t,z) = \mathbf{\widehat{U}}(z) + \mathbf{\widetilde{U}}(t, z)$, with the Floquet ansatz
\begin{equation*} 
  \mathbf{\widetilde{U}}(t, z) = e^{\lambda t} \mathbf{V}(z).
\end{equation*}
Here, $\lambda \in \mathbb{C}$ is the temporal eigenvalue and $\mathbf{V}(z) \in \mathbb{C}^{3}$ is the associated eigenfunction.
Linearization of system~\cref{eq:Tframe} about the stationary solution $\mathbf{\widehat{U}}$ leads to a second-order ODE for the eigenfunction $\mathbf{V}(z)$,
\begin{equation}
  \label{eq:eigenvalueProblem}
  \mathbf{V}_{zz} - \gamma \mathbf{V}_z + \mathbf{D_U f}(\mathbf{\widehat{U}}) \, \mathbf{V} = \lambda \, \mathbf{V},
\end{equation}
where $\mathbf{D_U f(\widehat{U})}$ is the Jacobian of $\mathbf{f(U)}$ evaluated at $\mathbf{\widehat{U}}$.
Solutions of this eigenvalue problem correspond to the spectrum of the linear operator $\mathcal{L} = \partial_{zz} - \gamma \, \partial_z + \mathbf{D_U f}(\mathbf{\widehat{U}}) $, acting on $\mathbf{V}(z)$. Spectra of linear operators separate into discrete and continuous parts; they are known as the point and essential spectra, respectively. It can be shown that the point spectrum of periodic TWs in reaction-diffusion models is always empty~\cite{sandstede2002stability}, so we will use `spectrum' to refer to the essential spectrum throughout this paper. A periodic TW solution $\mathbf{\widehat{U}}$ is said to be spectrally (or linearly) stable if the spectrum of the associated linear operator $\mathcal{L}$, which always contains the origin, lies otherwise entirely in the left half of the complex plane. Conversely, if the spectrum contains any eigenvalues with positive real part, then the corresponding periodic TW is linearly unstable.

\subsection{Set-up for computing the essential spectrum of periodic travelling waves}
In order to find the essential spectrum of the operator $\mathcal{L}$, we take into account the periodicity in $z$ of the underlying TW solution $\mathbf{\widehat{U}}$. The eigenfunction $\mathbf{V}(z)$ satisfies this same periodicity modulo a contraction or expansion associated with the (spatial) eigenvalues of $\mathcal{L}$.  We consider the eigenvalue problem~\cref{eq:eigenvalueProblem} with temporal eigenvalue $\lambda \in \mathbb{C}$ as a system of first-order ODEs,
\begin{equation}
  \label{eigenf}
  \left\{
    \begin{array}{rcl}
      \mathbf{V}'    &=& \mathbf{V}_z, \\
      \mathbf{V}_z' &=&   \gamma \mathbf{V}_z - \mathbf{D_U f}(\mathbf{\widehat{U}}) \, \mathbf{V} + \lambda \, \mathbf{V},
    \end{array} \right.
\end{equation}
subject to the boundary conditions
\begin{equation*}
  \left\{
    \begin{array}{lcl}
      \mathbf{V}(L)    &=& e^{\nu }  \, \mathbf{V}(0), \\
      \mathbf{V}_z(L) &=& e^{\nu }  \, \mathbf{V}_z(0),
    \end{array} \right.
\end{equation*}
where, $\nu \in \mathbb{C}$ is a spatial Floquet exponent measured with respect to the unit for space. Each eigenvalue $\lambda$ in the complex plane admits six complex Floquet exponents $\nu_1,\nu_2,...,\nu_6$. The essential spectrum is computed by assuming that no contraction or expansion is achieved in the spatial dimension. Hence, each Floquet exponent is restricted to the form $\nu = i \, \phi$, where the parameter $\phi \in \mathbb{R}$ represents the phase shift across one period of the TW. In other words, the boundary conditions are
\begin{equation}
  \label{BCs}
  \left\{
    \begin{array}{lcl}
      \mathbf{V}(L)    &=& e^{i \phi} \, \mathbf{V}(0), \\
      \mathbf{V}_z(L) &=& e^{i \phi} \, \mathbf{V}_z(0).
    \end{array} \right.
\end{equation}
In order to identify the eigenfunctions uniquely, we normalize and impose a phase condition as follows
\begin{equation}
  \label{eq:ICs}
  \left\{
    \begin{array}{l}
      {\displaystyle  \int_{0}^{L} \| \mathbf{V}(z) \|^2 \; dz = L }, \\[4mm]
      {\displaystyle  \int_{0}^{L} \textrm{Im} \left(\ \langle \mathbf{V}_{\rm old}(z) , \, \mathbf{V}(z)\, \rangle \ \right) \; dz = 0 }.
    \end{array} \right.
\end{equation}
Here, the brackets $\langle~,~\rangle$ denote the dot product and the subscript `old' denotes the solution from the previous (or initial) solution in the continuation step. The (essential) spectrum of a periodic TW solution $\mathbf{\widehat{U}}$ is the set of temporal eigenvalues $\lambda$, for which there exist spatial exponents $i \phi$ and associated eigenfunctions $V$ that satisfy the boundary value problem~\cref{eigenf}--\cref{eq:ICs}.

We encountered several challenges when computing the spectrum of periodic TWs in system~\cref{eq:MeanField} that are not necessarily specific to this example. The periodic TW solution will be similar to the heteroclinic cycle, especially when the wavelength is large.
A main challenge is that the alternating proximity of solution curves to the invariant subspaces increases numerical sensitivity for small values of the population variables and ultimately leads to numerical instability.
Since the population values are always non-negative,  we avoid this issue by writing the problem in logarithmic coordinates. To this end, we define $\mathbf{U}(t,z) = \exp[\mathbf{W}(t,z)]$ and consider the corresponding equation for $\mathbf{W}$ in the travelling frame:
\begin{equation}
  \label{eq:6ODEsLOG}
  \left\{
    \begin{array}{lcl}
      \mathbf{W}'    &=& \mathbf{W}_z, \\
      \mathbf{W}_z' &=& \gamma \mathbf{W}_z - \mathbf{g}(\mathbf{W}) - (\mathbf{W}_z)^2,
    \end{array} \right.
\end{equation}
 where $\mathbf{g}(\mathbf{W})$ satisfies $\mathbf{g}(\mathbf{W}(t, z)) = \exp[-\mathbf{W}(t, z)] \, \circ \, \mathbf{f}(\exp[\mathbf{W}(t, z)])$ and $\circ$ represents the Hadamard product (component-wise vector multiplication). Using a similar derivation as before, we linearize about the stationary solution $\mathbf{\widehat{W}}(z) = \log( \mathbf{\widehat{U}}(z) )$ and consider the perturbation in logarithmic coordinates, based on the same Floquet ansatz,
\begin{equation*} 
  \mathbf{W}(t,z) = \mathbf{\widehat{W}}(z) +  e^{\lambda t} \mathbf{V}(z),
\end{equation*}
which leads to the eigenvalue problem
\begin{equation*}
  \left\{
    \begin{array}{lcl}
      \mathbf{V}'    &=& \mathbf{V}_z, \\
      \mathbf{V}_z' &=& (\gamma  - 2 \mathbf{\widehat{W}}_z)  \,  \mathbf{V}_z - \mathbf{D_{\widehat{W}} g} \, \mathbf{V} + \lambda \, \mathbf{V},
    \end{array} \right.
\end{equation*}
with the same boundary and integral conditions~\cref{BCs} and~\cref{eq:ICs}. Here, the abbreviation $\mathbf{D_{\widehat{W}} g}$ represents the Jacobian $\mathbf{D_w g}$ of $\mathbf{g}(\mathbf{W})$ evaluated at $\mathbf{\widehat{W}}$.

A second major challenge, when the periodic TW solution for large wavelengths is similar to a heteroclinic cycle, is the fact that the solution also spends a relatively long time in the vicinity of each of the equilibria. Consequently, the solution variation over one period is extremely non-uniform, signalling that this is a very stiff problem. In particular, even though each eigenfunction is associated with Floquet exponents on the imaginary axis, there can still be enormous expansion and contraction pointwise along $\mathbf{V}$. We found it helpful to rescale $\mathbf{V}(z) \mapsto e^{-i \phi z} \, \mathbf{V}(z)$, so that the rescaled eigenfunction is periodic in $z$. With slight abuse of notation, we use the variable $\mathbf{V}$ for the rescaled version to obtain the following boundary value problem with periodic boundary conditions.
\begin{equation}
  \label{eigenfLOGperiodicBCs}
  \left\{
    \begin{array}{lcl}
      \mathbf{V}'    &=& \mathbf{V}_z - i \phi \, \mathbf{V}, \\[1mm]
      \mathbf{V}_z' &=&   (\gamma  - 2 \mathbf{\widehat{W}}_z) \, \mathbf{V}_z - \mathbf{D_{\widehat{W}} g} \, \mathbf{V} + \lambda \mathbf{V} - i \phi \, \mathbf{V}_z, \\[1mm]
      \mathbf{V}(L)     &=& \mathbf{V}(0), \\
      \mathbf{V}_z(L)  &=&  \mathbf{V}_z(0).
    \end{array} \right.
\end{equation}
System~\cref{eigenfLOGperiodicBCs} needs to be solved together with integral conditions~\cref{eq:ICs} and this computation is usually done in parallel with solving system~\cref{eq:6ODEsLOG} together with appropriate periodicity and phase conditions.

As a final remark, we note that the instructions given in~\cite{rademacher2007computing} suggest to find the initial data for the eigenfunction $\mathbf{V}(z)$ by way of a large matrix eigenvalue problem obtained from finite difference approximations of the discretized linear operator $\mathcal{L}$ based on a known periodic TW solution $\mathbf{\widehat{W}}$. In both papers~\cite{rademacher2007computing, sandstede2002stability}, the comment is made that $\lambda = 0$ is always contained in the spectrum and its associated eigenfunction is $\mathbf{\widehat{W}}_z$. Hence, it should be straightforward to start the continuation from $\lambda = 0$, with $\phi = 0$, periodic TW solution $\mathbf{\widehat{W}}$ and eigenfunction $\mathbf{\widehat{W}}_z$. More precisely, it suffices to consider the vector set $\{ \mathbf{\widehat{W}}, \mathbf{\widehat{W}}_z \}$ that solves system~\cref{eq:6ODEs}, and fix the associated eigenfunction $\{ \mathbf{V}, \mathbf{V}_z \}$ as $\mathbf{V} = \mathbf{\widehat{W}}_z$ and $\mathbf{V}_z = \mathbf{\widehat{W}}_{zz}$. 
When $\mathbf{V}$ and $\mathbf{V}_z$ are obtained from evaluating the right-hand side of system~\cref{eq:6ODEs}, there is no reason to assume that this eigenfunction has unit integral norm.
In practice, the eigenfunction will be scaled to satisfy the normalization condition as part of the corrections in the first continuation step, but convergence can be slow~\cite{rademacher2007computing}.
The tangent Floquet bundle can also be solved using a bifurcation approach, where the non-trivial eigenfunction $\{ \mathbf{V}, \mathbf{V}_z \}$ is computed as the solution branch that crosses the trivial solution $\mathbf{V} \equiv \mathbf{V}_z \equiv \mathbf{0}$ at a branching bifurcation that exists for $\phi = 0$; see~\cite{kr-Lin2008, om-siads2010} for more details. The advantage of this bifurcation approach is that the non-trivial eigenfunction is first detected as a solution with zero norm emanating from the branching bifurcation point; the branching step grows the eigenfunction until it satisfies the unit integral norm in condition~\cref{eq:ICs}. 
Detecting the desired Floquet exponent $\nu$ can sometimes be tricky for stiff problems.
However, we know that the branch point lies at $\phi = 0$ and are able to skip this detection step as follows. We start the branching step at $\phi = 0$ from the trivial branch but with an initial eigenfunction $\{ \mathbf{V}, \mathbf{V}_z \}$ that is an arbitrary constant function with (small) nonzero norm; this nonzero norm ensures that {\sc Auto} corrects and converges to the true tangent Floquet bundle associated with $\phi=0$, despite the lack of branching information that would normally be encoded in the detected branch point.

\subsection{Geometry of essential spectra: An example}
%
\begin{figure}[t]  
  \centering
  \includegraphics{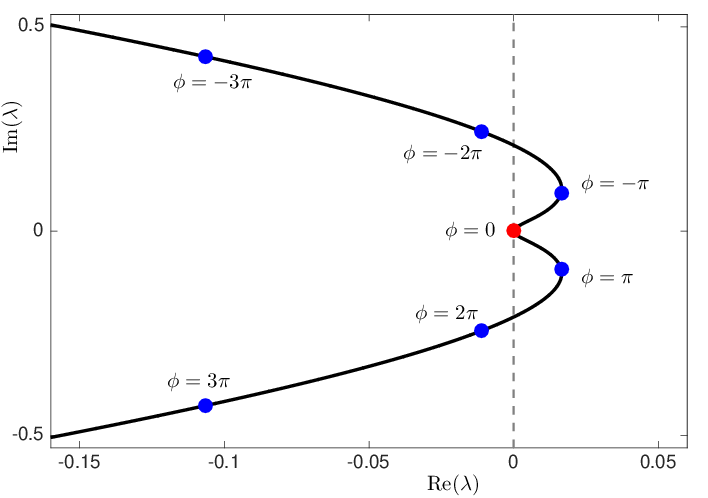}
  \caption{\label{fig:spec}
    Essential spectrum (black curve) of the periodic TW for system~\cref{eq:MeanField} with $\sigma=3.2$, $\zeta=1$, $L=20$ and corresponding wavespeed $\gamma \approx 1.69505$. The coloured dots indicate a selection of phases $\phi$.}
\end{figure}
%
\Cref{fig:spec} shows the spectrum of a periodic TW at $\sigma=3.2$, $\zeta=1$ and $L=20$  (with corresponding wavespeed $\gamma \approx 1.69505$) as an example. Vertical and horizontal axes correspond to the real and imaginary parts of the temporal eigenvalue $\lambda$, respectively. 
The essential spectrum is computed by continuation of solutions to the boundary value problem~\cref{eq:ICs}--\cref{eigenfLOGperiodicBCs}, starting from the origin $\lambda = 0$ (red dot) with $\phi = 0$ and initial data obtained by substituting the known periodic TW solution $\mathbf{\widehat{W}}$ into the right-hand side of system~\cref{eq:6ODEsLOG}; here, both $\lambda$ and $\phi$ are allowed to vary. The black curve in \cref{fig:spec} is the spectrum of the periodic TW and the blue dots are selected $\lambda$-values that correspond to various phases $\phi$. Note that the spectrum is symmetric about the real axis due to the translational invariance of the linear operator $\mathcal{L}$. 

\begin{figure}[t] 
  \centering
  \includegraphics{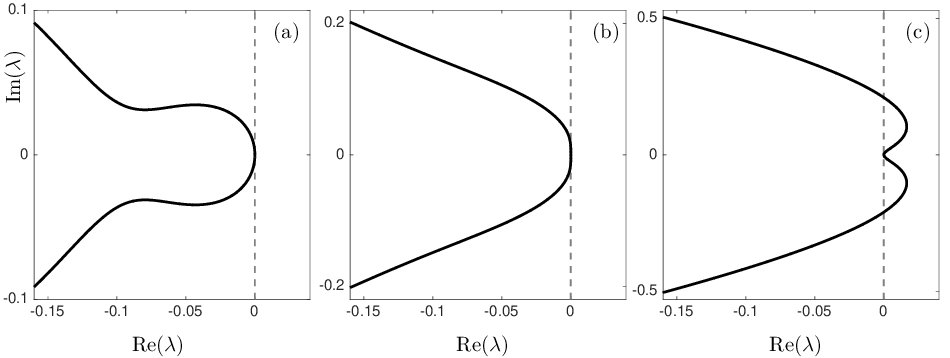}
  \caption{\label{fig:EckMech}
    Illustration of an Eckhaus-type change in stability of periodic TWs of system~\cref{eq:MeanField} with $\sigma=3.2$ and $\zeta=1$ as the wavelength $L$ varies. Shown are the essential spectrum before~(a) with $L = 40$ ($\gamma \approx 1.90633$), approximately at~(b) with $L \approx 29.9950$ ($\gamma \approx 1.83521$), and after the bifurcation~(c) with $L = 20$ ($\gamma \approx 1.69505$), which is the same value as used in \cref{fig:spec}.}
\end{figure}
%
The spectrum of the periodic TW for $\sigma=3.2$, $\zeta=1$ and $L=20$ extends into the right half of the complex plane, which indicates that the periodic TW is linearly unstable for this choice of parameters. \Cref{fig:EckMech} shows three spectra of periodic TWs in~\cref{eq:MeanField} with $\sigma=3.2$ and $\zeta=1$. Here, we vary the wavelength $L$ and, consequently, the wavespeed $\gamma$. Panel~(a) shows the spectrum of the periodic TW with $L = 40$ and $\gamma \approx 1.90633$. Observe that, apart from the origin, the spectrum for this wavelength lies entirely in the left half of the complex plane. Hence, the corresponding periodic TW is linearly stable. Panel~(b) illustrates the onset of an instability, approximately at  $L = 29.9950$ ($\gamma \approx 1.83521$). The tangency at the origin made by the  (black) curve of eigenvalues $\lambda$ appears to be of a higher order than quadratic, suggesting that the curvature of the spectrum is (almost) zero; this type of instability is known as an Eckhaus instability~\cite{Eckhaus}. Panel~(c) shows the same spectrum for $L = 20$ as in \cref{fig:spec}, which curves in the opposite direction into the right half of the complex plane; the corresponding periodic TW is now unstable.

\subsection{Changes in stability: the Eckhaus bifurcation}
\label{sec:Eckhaus}
There are two common types of instabilities that can occur for the essential spectrum of periodic TWs~\cite{rademacher2007computing}. The Eckhaus instability~\cite{Eckhaus} illustrated in \cref{fig:EckMech} is also known as sideband instability and occurs when the curvature of the spectrum changes sign at the origin ($\lambda = 0$); this is the only type of instability that we find in system~\cref{eq:MeanField}. Hopf instability occurs when a pair of complex eigenvalues on the spectrum crosses the imaginary axis away from the origin; our numerical explorations did not indicate the existence of a Hopf instability for the periodic TWs in system~\cref{eq:MeanField}.

Loss of stability via an Eckhaus instability can be computed numerically as a zero of the second derivative $\frac{d^2 }{d \nu^2}\textrm{Re}(\lambda) \big|_{\lambda=\phi=0}$ with respect to $\nu = i \, \phi$ of the real part of the eigenvalue $\lambda$ evaluated at the origin; indeed, the curve of eigenvalues in the complex plane is parametrized by $\nu = i \, \phi$. The onset for $\sigma=3.2$ and $\zeta=1$ can readily be detected this way and we find that an Eckhaus instability occurs when $L \approx 29.9950$. A subsequent two-parameter continuation in $L$ and $\zeta$ determines the boundary of the region of stability in the $(\zeta, L)$-plane, by imposing the condition $\frac{d^2 }{d \nu^2}\textrm{Re}(\lambda) \big|_{\lambda=\phi=0} = 0$.

The formulation of the corresponding boundary value problem is described in~\cite{rademacher2007computing}, where the required second derivative is found via implicit differentiation of the eigenvalue problem with respect to $\nu= i \phi$. We adapt the set-up to our system~\cref{eigenfLOGperiodicBCs}, which is written in logarithmic coordinates. We use the notation $\mathbf{V}_\phi$, $\mathbf{V}_{\phi \phi}$ and $\mathbf{V}_{z \phi}$, $\mathbf{V}_{z \phi \phi}$ for the first and second derivatives with respect to $i \phi$ of $\mathbf{V}$ and $\mathbf{V}_z$ at $\lambda = \phi = 0$, respectively. 
Then the boundary value problem is extended by the following system of equations,
\begin{equation}
    \left\{
      \begin{array}{rcl}
        (\mathbf{V}_{\phi})'           &=& \mathbf{V}_{z \phi} - \mathbf{V}, \\[1mm]
        \left( \mathbf{V}_{z \phi} \right)'      &=& (\gamma - 2 \mathbf{\widehat{W}}_z) \, \mathbf{V}_{z \phi \phantom{\phi}} -
              \mathbf{D_{\widehat{W}} g} \, \mathbf{V}_{\phi \phantom{\phi}} + \lambda_{\phi \phantom{\phi}} \mathbf{V} -
              \mathbf{V}_z \\
        (\mathbf{V}_{\phi \phi})'      &=& \mathbf{V}_{z \phi \phi} - 2 \mathbf{V}_{\phi}, \\[1mm]
        (\mathbf{V}_{z \phi \phi})' &=& (\gamma - 2 \mathbf{\widehat{W}}_z) \, \mathbf{V}_{z \phi \phi} -
             \mathbf{D_{\widehat{W}} g} \, \mathbf{V}_{\phi \phi} + \lambda_{\phi \phi} \, \mathbf{V}  + 2 \lambda_{\phi} \, \mathbf{V}_{\phi} -
                                                                                                                                                                                               2 \, \mathbf{V}_{z \phi},
      \end{array} \right.
\end{equation}
subject to the periodic boundary conditions
\begin{equation}
  \left\{
    \begin{array}{lcl}
      \mathbf{V}_{\phi} (L)          &=& \mathbf{V}_{\phi} (0), \\
      \mathbf{V}_{z \phi} (L)       &=& \mathbf{V}_{z \phi} (0), \\
      \mathbf{V}_{\phi \phi} (L)    &=& \mathbf{V}_{\phi \phi} (0), \\
      \mathbf{V}_{z \phi \phi} (L) &=& \mathbf{V}_{z \phi \phi} (0),
    \end{array} \right.
\end{equation}
\noindent
and integral conditions 
\begin{equation}
  \label{ICs}
  \left\{
    \begin{array}{l}
      {\displaystyle   \int_{0}^{L} \langle\ \mathbf{V}(z),\, \mathbf{V}_{\phi} (z) \, \rangle \; dz = 0 }, \\[4mm]
      {\displaystyle  \int_{0}^{L} \langle\ \mathbf{V}(z),\, \mathbf{V}_{\phi \phi} (z) \, \rangle \; dz = 0 }.
    \end{array} \right.
\end{equation}
%

\begin{figure}[t] 
  \centering
  \includegraphics{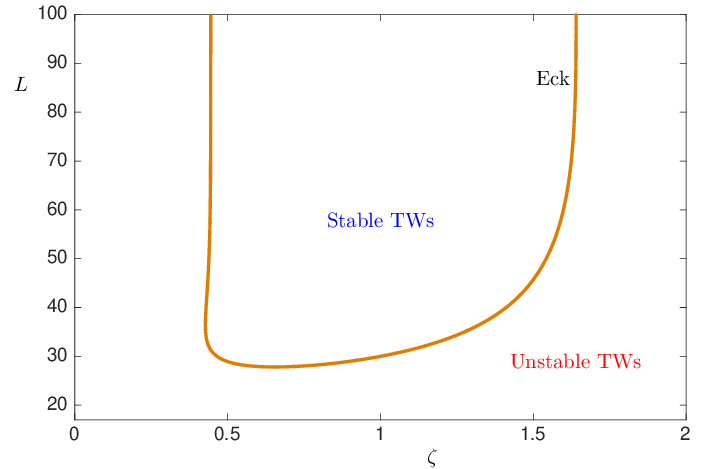}
  \caption{\label{fig:locus}
    Locus of Eckhaus instability (Eck) of periodic TWs of system~\cref{eq:MeanField} in the ($\zeta,L$)-parameter plane, with $\sigma=3.2$.}
\end{figure}
%
\Cref{fig:locus} shows the curve of Eckhaus instability, denoted Eck (orange), in the $(\zeta, L)$-parameter plane, which marks the instability onset of periodic TWs of system~\cref{eq:MeanField} with $\sigma = 3.2$. In this projection, there exists a single periodic TW at every point of the parameter plane. The convex region bounded by the curve Eck is called a Busse balloon \cite{Busse67} and it exclusively contains all stable periodic TWs.
In the region outside the Busse balloon, periodic TWs are linearly unstable. We observe from \cref{fig:locus}  that small-wavelength periodic TWs will always be unstable. In contrast, large-wavelength TWs are stable within a finite interval of $\zeta$.
Note that this finding is rather counter-intuitive to the heuristic explanation in the introduction of the fronts catching up quickly enough to squash any perturbation of the wave.

\subsection{Busse balloon and the bifurcation diagram for different values of $\sigma$}
%
\begin{figure}[t]
  \centering
  \includegraphics{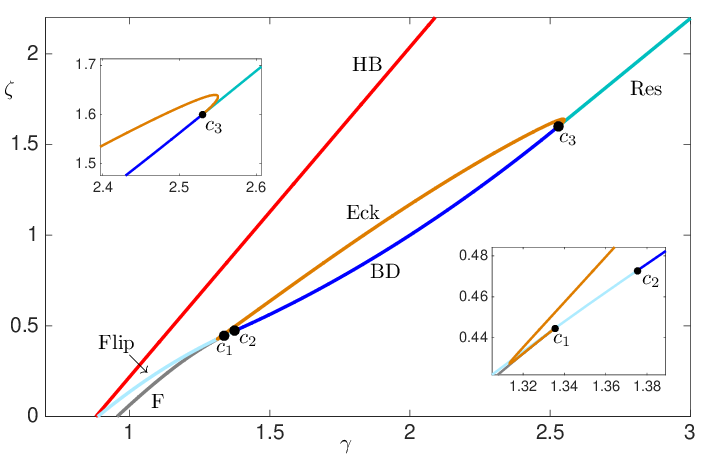}
  \caption{\label{fig:EckBif1}
    Bifurcation diagram with the curve Eck of the Eckhaus instability (orange) for system~\cref{eq:MeanField} with $\sigma = 3.2$. The insets illustrate how the Eckhaus instability curve ends at the codimension-two points $c_1$ and $c_3$ (black dots). See \cref{fig:ODEbif} for details on the other curves and labels.}
\end{figure}
%
In this section, we investigate how the stability region (Busse balloon) depends on $\sigma$.
The Busse balloon corresponds to the region bounded by the curves Eck, BD, and the segment between $c_1$ and $c_3$ of the curve Flip.
 \Cref{fig:EckBif1} shows the curve Eck of the Eckhaus instability (orange) superimposed on the bifurcation diagram of the travelling-frame ODE system~\cref{eq:MeanField} for $\sigma = 3.2$; compare with \cref{fig:ODEbif}. As before, we show the loci of Hopf bifurcation (HB, red), fold of periodic orbits (F, grey), and heteroclinic bifurcations of flip (Flip, light blue), Belyakov--Devaney (BD, dark blue) and resonance types (Res, turquoise). The two insets show enlargements in the vicinity of the codimension-two points. We find that the curve Eck begins and terminates at the codimension-two points $c_1$ and $c_3$, where it is tangent to the curves Flip and Res.
\begin{figure}[t]
  \centering
  \includegraphics[scale=0.974]{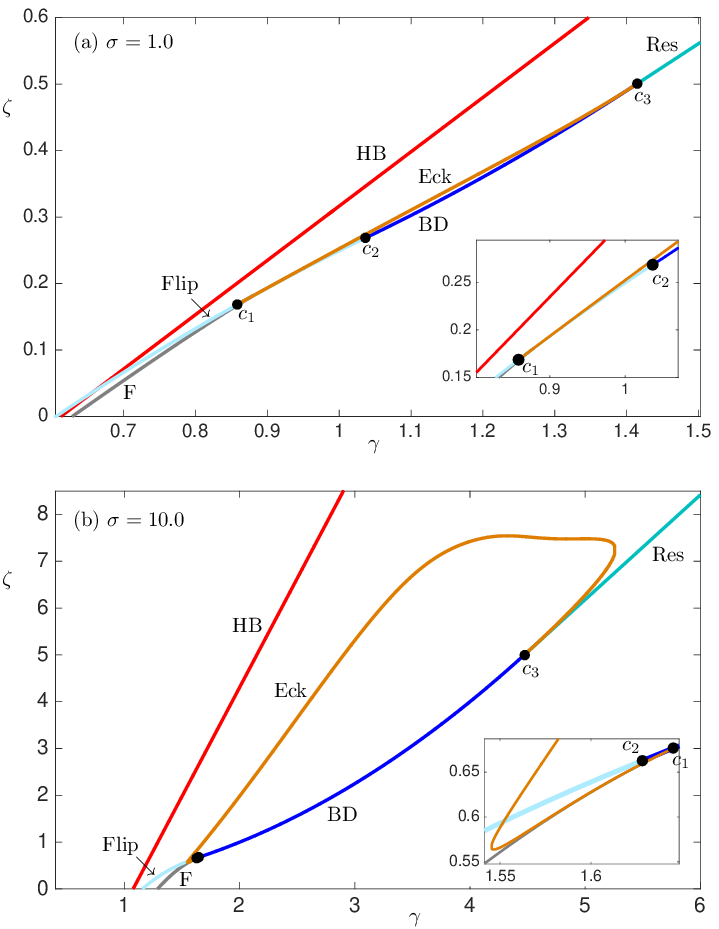}
  \caption{\label{fig:EckBif2}
    Bifurcation diagram with Eckhaus instability curve for system~\cref{eq:MeanField} with $\sigma = 1.0$~(a) and $\sigma=10.0$~(b). Colours and labels are as in \cref{fig:EckBif1}.}
\end{figure}
%
We repeat the computation of the Eckhaus instability for smaller and larger values of $\sigma > 0$. \Cref{fig:EckBif2} shows the bifurcation diagrams of system~\cref{eq:6ODEsLOG} in panels~(a) and~(b) for $\sigma = 1.0$ and $\sigma = 10.0$, respectively; compare with \cref{fig:EckBif1}. Panel~(a) shows that, for $\sigma = 1.0$, the Busse balloon shrinks down to a thin region in the $(\gamma, \zeta)$-plane that is bounded by the curves Eck (orange), BD (dark blue) and Flip (light blue). As before, the end points of the curve Eck are the codimension-two points $c_1$ and $c_3$. For smaller values of $\sigma > 0$, the Busse balloon becomes smaller and smaller and  the Eckhaus curve lies flat against the curve of heteroclinic bifurcations as $\sigma \to 0$. 
The region of existence of periodic orbits also becomes smaller when decreasing $\sigma$, and we observe that the bifurcation curve HB collides with the curves F, Flip, BD and Res as $\sigma \to 0$.

In \cref{fig:EckBif2}(b), the stable regime for $\sigma = 10.0$ extends along a wider region in the $(\gamma, \zeta)$-plane. The curve Eck still terminates at $c_1$ and $c_3$, but $c_1$, defined as the point where the curve F terminates, now lies to the right of $c_2$ and both F and Eck terminate on the curve BD. The details of how this transition occurs are left for future work.

\begin{figure}[t]
  \centering
  \includegraphics[scale=0.974]{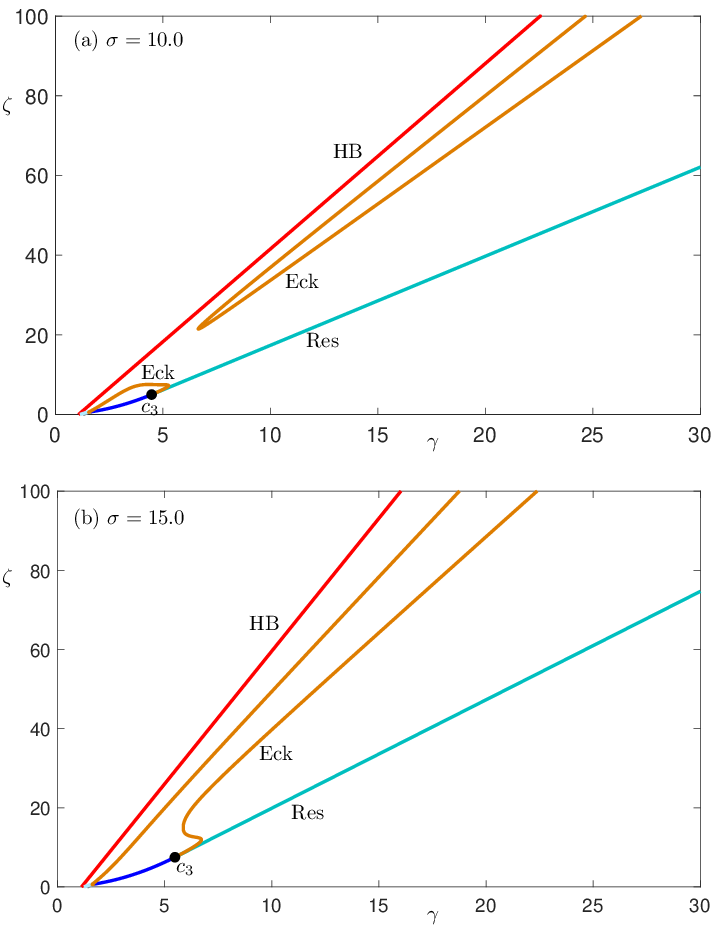}
  \caption{\label{fig:EckBif3}
    Bifurcation diagram with Eckhaus instability curve for system~\cref{eq:MeanField} with $\sigma = 10.0$~(a) and $\sigma = 15.0$~(b) for larger ranges of $\gamma$ and $\zeta$. Colours and labels are as in \cref{fig:EckBif1}.}
\end{figure}
%
We detected another regime of stable periodic TWs that is secondary to the Busse balloon. \Cref{fig:EckBif3} shows the bifurcation diagrams of system~\cref{eq:6ODEsLOG} for larger ranges of $\gamma$ and $\zeta$. Here, $\sigma = 10.0$ and $\sigma = 15.0$ in panels~(a) and~(b), respectively. For $\sigma = 10$, as shown in panel~(a), we find a second branch of Eckhaus bifurcation that appears to extend to infinity and forms a convex region that constitutes a second stable regime. We were unable to detect such a second stable region for $\sigma = 3.2$ and we suspect that this region exists only for very large values of $\zeta$ when $\sigma$ is relatively small. For $\sigma \approx 11.5239$, the two stable regions merge into a single stable region.
\Cref{fig:EckBif3}(b) shows that there is a single stable region for $\sigma = 15.0$ that extends to infinity. For larger values of $\sigma$, we observe that the stable region extends to cover an increasingly larger region in the $(\gamma,\zeta)$-plane.

\section{Comparison to PDE simuations}
\label{sec:simulations}
In this section, we compare the stability analysis to direct simulations of one-dimensional TWs and two-dimensional SWs for the PDE system~\cref{eq:MeanField}.
Throughout, integration of the PDE system~\cref{eq:MeanField} is carried out using a second-order exponential time differencing method for integrating stiff systems \cite{Cox2002} with periodic boundary conditions, i.e., $\mathbf{U}(t,L)=\mathbf{U}(t,0)$ and $\mathbf{U}_x(t,L)=\mathbf{U}_x(t,0)$.

The stability analysis predicted by the computation of the essential spectrum corresponds to the stability of periodic TWs that extend along an infinitely large domain $-\infty<x<\infty$.
Of course, this is unfeasible in practice.
Therefore, in order to have a good agreement with this analysis, we carry out simulations of five copies of periodic TWs along a large bounded domain of size $5 \times L$, where $L$ is the wavelength of the periodic TW.
When the number of copies is increased, we obtain qualitatively similar results but the computation time is significantly longer. 
Throughout this section, the initial condition is a train of five periodic TWs with wavelength $L$, which is obtained from the numerical continuation of periodic solutions of system~\cref{eq:MeanField}.
%
\begin{figure}[t]
  \centering
  \includegraphics{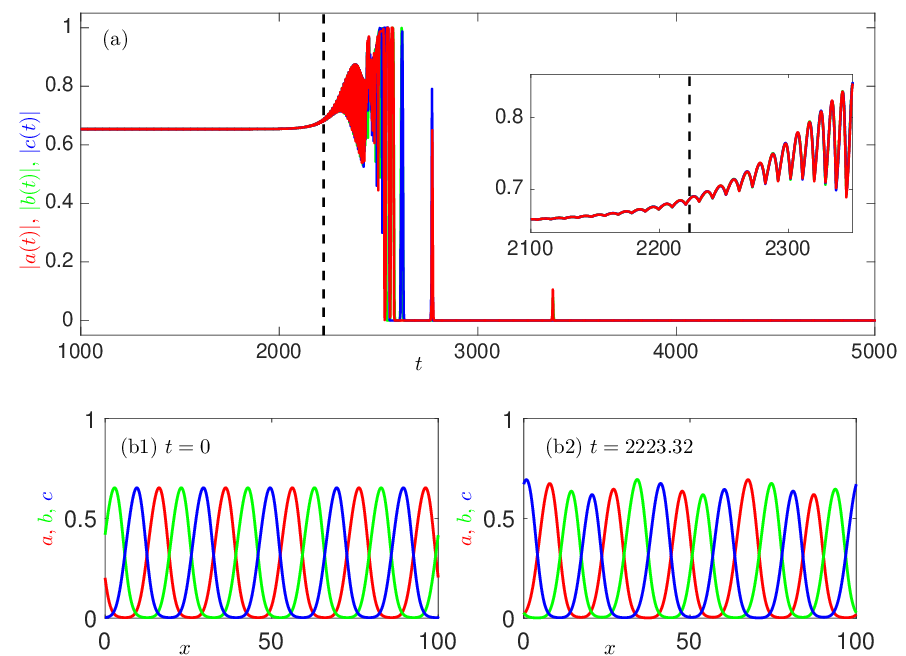}
  \caption{\label{fig:instabilityMech}
    Illustration of the instability mechanism for periodic TWs in system~\cref{eq:MeanField} with $\sigma=3.2$, $\zeta=1.0$. Panel~(a) shows the amplitudes of the unstable periodic TW for $a$ (red), $b$ (green) and $c$ (blue), as a function of time; the inset is an enlargement. The dashed black line indicates the moment at which the amplitude of the TW increases by $5\%$. Panels~(b1) and~(b2) show the the distributions of the three populations for the initial condition ($t=0$) and at the heuristic moment of instability ($t=2223.32$), respectively.}
\end{figure}

\subsection{Instability mechanism of periodic TWs in system~\cref{eq:MeanField}}
We find that unstable periodic TWs in system~\cref{eq:MeanField} deform in a particular way as follows.
\Cref{fig:instabilityMech} demonstrates the instability mechanism for periodic TWs in system~\cref{eq:MeanField} with $\sigma=3.2$, $\zeta=1$ and $L=20$.
Panel (a) shows the amplitudes of the three populations as a function of time; the inset is an enlargement highlighting the onset of the instability.
Here, the amplitude of a variable refers to the difference between the maximum and minimum of that variable over the entire domain.
The initial condition shown in \cref{fig:instabilityMech}(b1) is obtained from a periodic solution to the travelling-frame system~\cref{eq:6ODEsLOG}; its essential spectrum is shown in \cref{fig:EckMech}(c).
Note that all three variables appear to have approximately fixed amplitudes for a certain amount of time, but at $t \approx 2000$ the amplitudes of the wave begin to oscillate in time and an irregular spatiotemporal behaviour emerges.
Eventually, the amplitudes of the waves decay to zero and each population acquires a solution that is homogeneous in space.
The acquired homogeneous solutions are then governed by the following ODE
\begin{equation} 
\mathbf{U}_t=\mathbf{f(U)},
\end{equation}
and approach a heteroclinic cycle in time.

We observe that this instability mechanism is characteristic for all unstable periodic TWs in system~\cref{eq:MeanField}. 
Therefore, we use the following criterion for the onset of instability of periodic TWs in system~\cref{eq:MeanField}.
We define the moment at which the amplitude of one of the three populations changes by $5 \%$ as a heuristic criterion for observing the onset of instability; see \cref{fig:instabilityMech}(b2).
The vertical dashed line at $t = 2223.32$ in \cref{fig:instabilityMech}(a) corresponds to this moment for the simulation; see also the enlargement in the inset.

\begin{figure}[t]
  \centering
  \includegraphics{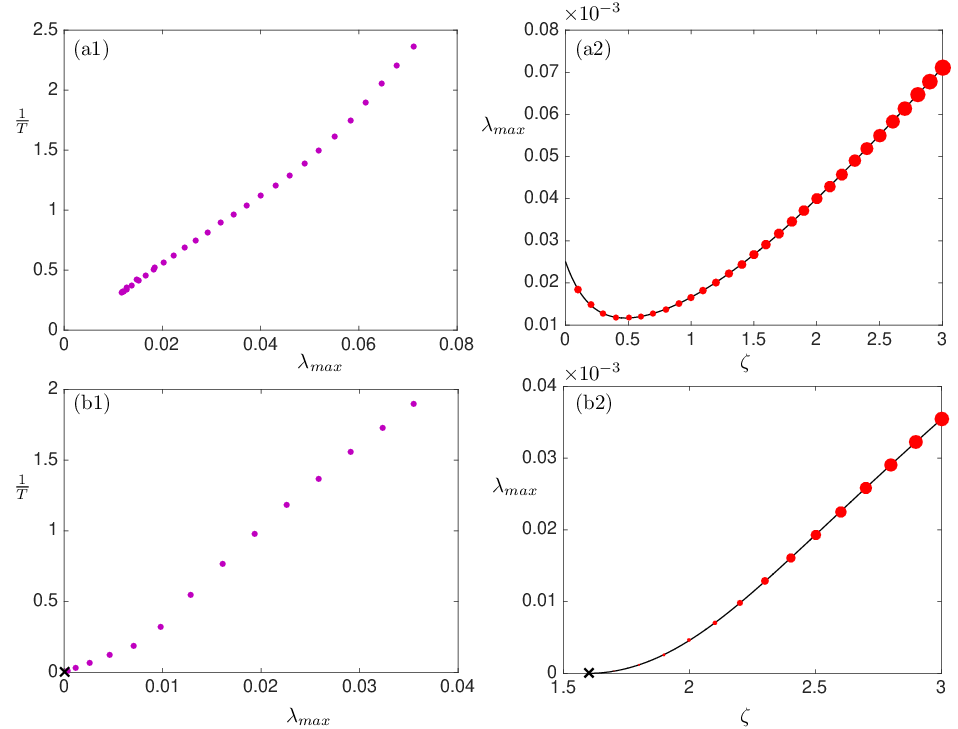}
  \caption{\label{fig:eigenvalues}
    Comparison between PDE simulations of the one-dimensional periodic TWs and the respective spectral stability analysis of system~\cref{eq:MeanField} with $\sigma=3.2$. Rows~(a) and~(b) correspond to periodic TWs with wavelengths $L=20$ and $L=60$, repsectively. The left column shows $\frac{1}{T}$ versus the leading temporal eigenvalue $\lambda_{max}$, where $T$ is the minimal integration time needed for the instabilities to be observable.
In the right column, the black curve is the dispersion relation between $\lambda_{max}$ and $\zeta$, and the red circles represent the PDE simulations for parameter values at their centres. The radii of the red circles are proportional to $\frac{1}{T}$. The black cross corresponds to a parameter value at which the instability was not observed in the PDE simulations.}
\end{figure}

\subsection{Relation between instabilities in the simulated PDE system and the computed eigenvalues}
PDE simulations of unstable periodic TWs may take a very long time before the instabilities become observable.
Hence, we perform further spectral analysis for the unstable region (the region outside the Busse balloon).
Namely, we compute the maximal growth rate associated with the unstable TWs, that is, the real part of the leading (rightmost) temporal eigenvalue of the spectrum $\lambda_{max}:=\max\{$Re($\lambda$)$\} $.
The maximal eigenvalue $\lambda_{max}$ of unstable periodic TWs represents the rate at which the perturbations grow causing the destabilization of the waves \cite{Siteur2014}.
Therefore, we compare $\lambda_{max}$ to the growth rates of the unstable waves in the PDE simulations.

\Cref{fig:eigenvalues} shows a comparison between the computed maximal growth rate $\lambda_{max}$ obtained from equations \cref{eigenf}--\cref{ICs} and the integration time $T$ needed for the periodic TWs to destabilize in the PDE simulations according to our heuristic criterion.
In rows (a) and (b), we fix $L=20$ and $L=60$, respectively, and allow $\zeta$ to vary.
In the left column of \cref{fig:eigenvalues}, we plot the growth rate of the unstable periodic TWs measured by $\frac{1}{T}$ against the computed maximal growth rate $\lambda_{max}$. When $\lambda_{max}$ is sufficiently small,
the relation between the two quantities appears to be almost linear.
For $L=60$, we performed a least-squares fit to the log-transformed data for $\lambda_{max} < 0.01$, and found the approximate relationship
\begin{equation} \label{eq:LinearFormula}
 \log\left( \frac{1}{T} \right) \approx  1.049  \log(\lambda_{max})-3.364.
 \end{equation}
 The black crosses in panels (b1) and (b2) correspond to a point that lies just outside the stability
region shown in \cref{fig:locus}. However, the corresponding train of periodic TWs does not
show any sign of instability, despite integrating for the very long time of $T=3.0 \times 10^{5}$.
Indeed the computed growth rate $\lambda_{max}=8.51062 \times 10^{-8}$ suggest that an integration time $T>7.539 \times 10^8$ is needed before the destabilization can be observed.

In the right column of \cref{fig:eigenvalues}, we plot the dispersion relation (black curve) between the maximal eigenvalue $\lambda_{max}$ and $\zeta$, computed using the continuation software {\sc Auto}, again for $L=20$ (a2) and $L=60$ (b2).
The red circles on top of the dispersion relation are representatives of the growth rate $\frac{1}{T}$ in the PDE simulations.
More precisely, the radii of the circles are linearly proportional to $\frac{1}{T}$.
Note that the larger the maximal eigenvalue $\lambda_{max}$, the larger the corresponding data circles  (i.e., the faster the corresponding periodic TWs destabilize in the PDE simulations).

\subsection{Belts of instability}
We now identify different subregions in the ($\zeta,L$)-parameter plane based on the maximal growth rate $\lambda_{max}$; we call these subregions \emph{belts of instability}.
The idea is to fix the real part of the leading temporal eigenvalue $\lambda_{max}$ of the spectrum of an unstable TW and continue it as a contour curve in the two-parameter plane.

\begin{figure}[t]
  \centering
  \includegraphics{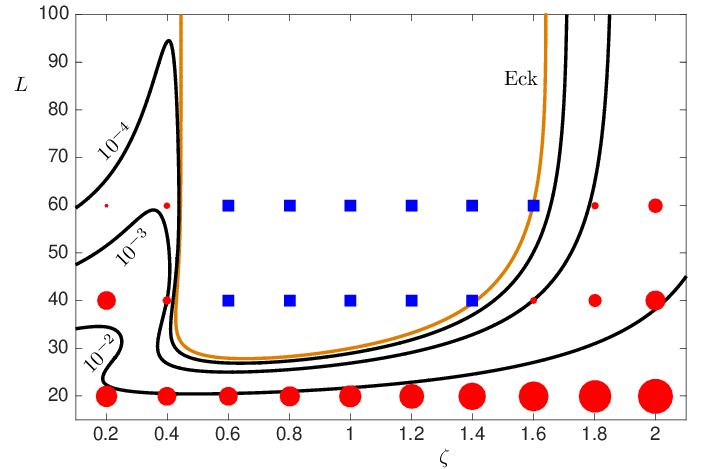}
  \caption{\label{fig:comparison}
    Comparison between PDE simulations of the one-dimensional periodic TWs and the respective spectral stability analysis of system~\cref{eq:MeanField} with $\sigma=3.2$. The orange curve is the curve of Eckhaus instability (Eck).
The black curves are contours for the leading temporal eigenvalues $\lambda_{max}= 10^{-2}$, $\lambda_{max}= 10^{-3}$, and $\lambda_{max}= 10^{-4}$  for the corresponding periodic TWs. The blue squares and red circles represent stable and unstable periodic TWs, respectively. The radii of the red circles are inversely proportional to the minimal integration time needed for the instabilities to be observable.}
\end{figure}

\Cref{fig:comparison} shows the unstable region (outside the Busse balloon) subdivided into four quantitative subregions, which are bounded by contours (black curves) of fixed leading eigenvalues $\lambda_{max}$.
For instance, the black curve labeled $10^{-2}$ is obtained by continuing the contour for the rightmost temporal eigenvalue $\lambda_{max}=10^{-2}$; see also the spectrum shown in \cref{fig:EckMech}.
One would expect that perturbations of the periodic TWs grow faster in the belts of instability with larger eigenvalues $\lambda_{max}$.
For example, the region bounded in between the curves labeled $10^{-3}$ and $10^{-2}$ contains periodic TWs that are `more unstable' than those that lie in between the curves labeled $10^{-4}$ and $10^{-3}$.

%
Also shown in \cref{fig:comparison} is the PDE simulation data superimposed on the different stability and instability regions.
The blue squares indicate that the simulated periodic TWs remain stable for the entire total integration time $T=3.0 \times 10^{5}$. 
The radii of the red circles are proportional to the growth rate $\frac{1}{T}$, where $T$ is the integration time needed to achieve our criterion of a $5 \%$ amplitude increase for the periodic TWs.
The blue square at $(\zeta, L)=(1.6,60)$ lies just outside the stable region and corresponds to the black cross in \cref{fig:eigenvalues}(b2).
We again find that the growth rate in the PDE simulations, measured by $\frac{1}{T}$, increases proportionally with the maximal eigenvalue $\lambda_{max}$.
For instance, note that periodic TWs at parameter values in belts with smaller eigenvalues  $\lambda_{max}$ take longer times to destabilize (the red circles are smaller).
The upshot is that the maximal growth rate $\lambda_{max}$ obtained from the spectral analysis is a very good quantitative indicator of the instability of the periodic TWs.

\subsection{Comparison to stability of spiral waves}
%
\begin{figure}[t]
  \centering
  {\includegraphics[scale=1.0]{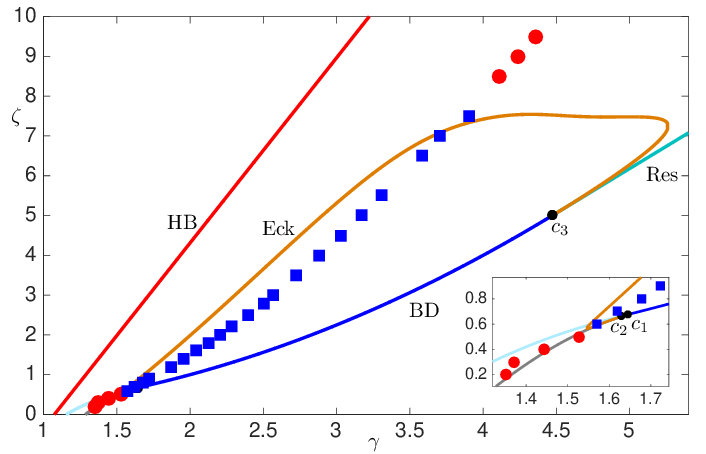}}
  \caption{\label{fig:spiralStability}
    Comparison between PDE simulations of the two-dimensional spiral waves and the spectral stability analysis of the asymptotic periodic TWs of system~\cref{eq:MeanField} with $\sigma=10.0$. The blue squares and red circles represent stable and unstable spiral waves, respectively. Colours and labels of bifurcation curves are as in \cref{fig:EckBif1}. The inset is an enlargement.}
\end{figure}
%
We illustrated that solutions of spiral waves converge asymptotically to a train of periodic TWs as they move away from the core; see \cref{fig:SWs}.
Sandstede and Scheel~\cite{sandstede2000absolute} have shown that the maximal eigenvalue of the spectrum of spiral waves coincides with that of the asymptotic periodic TW.
In other words, spiral waves and associated periodic TWs have the same stability boundary.
Therefore, we perform PDE simulations of the two-dimensional spiral waves in a very large box ($2000 \times 2000$) and compare it against the stability of the associated TWs. Our initial conditions resulted in four spirals centred in the middle, edges and corners of the box, but we extrapolated the central spiral to fill the box using polar coordinates centred on the core and Fourier transforms of the angular dependence of $\mathbf{U}$. This extrapolation process results in perturbations to the core, and the instability or stability of the spirals was judged by whether these perturbations grew or decayed, respectively, as they moved outwards. See the supplementary data link \url{https://github.com/CrisHasan/TW_Nonlinearity2021} for three examples.

\Cref{fig:spiralStability} shows the two-dimensional spiral wave simulation data superimposed on the bifurcation diagram for the one-dimensional periodic TWs in system~\cref{eq:MeanField} with $\sigma=10.0$.
Blue squares indicate stability and red circles indicate instability.
We note that the asymptotic TWs admit relatively short wavelengths $L \in (16.19,20.53)$ and are not particularly close to the heteroclinic bifurcations.
The red circles in \cref{fig:spiralStability} lie outside the stable Busse balloon and the blue squares lie inside it except for one data point at $(\gamma,\zeta)=(3.9023,7.5)$.
Since the curve Eck corresponds to the stability of the asymptotic periodic TWs in an infinitely unbounded domain, this discrepancy is likely due to the finite size of the domain in the simulations. For example, we performed a simulation at $(\gamma,\zeta)=(4.1082,8.5)$ in the smaller box of size $1000 \times 1000$, and found that the spiral wave then appears to be stable for the entire integration time, although, it is convectively unstable in a 2000x2000 box (see supplementary Material \url{https://github.com/CrisHasan/TW_Nonlinearity2021}).
Overall, we find that the stability analysis of asymptotic periodic TWs matches the stability of the associated spiral waves, as expected.

\section{Conclusions and discussion}
\label{sec:discussion}
We investigated the stability of periodic TWs in a spatially-extended May--Leonard system, a phenomenological model that describes the local dynamics of cyclic interactions of three competing populations.
The spatiotemporal behaviour in this model is intricate and not well understood.
In this study, we performed a linear stability analysis of periodic TWs and examined it against the PDE simulation in the laboratory frame. 
In particular, we computed the onset of Eckhaus instabilities and located the parameter regimes in which the periodic TWs are stable.
We also identified different unstable subregions (belts of instability) that quantitatively vary with respect to the underlying growth rate.

Numerical evidence from previous work \cite{Gani2016, sherratt2013numerical, Siero2015} on periodic TWs, or periodic pulse trains, generated in systems with an underlying homoclinic bifurcation (solitary pulse), shows that the Eckhaus instability curve terminates at codimension-two points that involve a homoclinic bifurcation.
In the context of heteroclinic-induced periodic travelling waves, namely system~\cref{eq:MeanField}, we found that the Eckhaus instability curve terminates at two codimension-two points $c_1$ and $c_3$, both of which involve heteroclinic bifurcations.
Robust heteroclinic cycles in system~\cref{eq:MeanField} are formed by concatenating three travelling fronts with unstable essential spectrum.
We have also performed preliminary calculations for the absolute spectrum of these travelling fronts and found that the onset of the absolute spectrum of the travelling fronts changes at the codimension-two points $c_2$ and $c_3$.
In particular, the largest eigenvalue of the absolute spectrum of the fronts is zero along the curve BD and less than zero along the curves Res and Flip.
In the $(\gamma,\zeta)$-parameter plane, near heteroclinic bifurcations, we find that the travelling fronts below and above the curve $\zeta=\gamma^2/4$  are absolutely stable and unstable, respectively.
We remark that Belyakov--Devaney-type heteroclinic bifurcation is described by the algebraic relation $\zeta=\gamma^2/4$.
This suggests that the stability of periodic TWs (organized by $c_1$ and $c_3$) as well as absolute stability of the connecting travelling fronts (organized by $c_2$ and $c_3$) are determined by degenerate heteroclinic bifurcations.
Since the organising codimension-two points are different, there are no obvious conclusions that can be drawn with respect to the relationship between stability of large-wavelength periodic TWs and the absolute stability of the travelling fronts forming the heteroclinic cycles.

The mechanism underlying the instability of TWs may take different forms.
We found that unstable periodic TWs in system~\cref{eq:MeanField} deform in a particular way that causes their wave amplitudes to oscillate in time. 
We introduced a heuristic criterion for the instability of periodic TWs and used it as a benchmark to measure the growth rate in the laboratory frame. 
Comparing the growth rate in the laboratory frame against the largest temporal eigenvalue $\lambda_{max}$ suggests that the rightmost eigenvalue of the essential spectrum captures the long term spatiotemporal behaviour of unstable periodic TWs in system~\cref{eq:MeanField}.

Spiral waves in the two-dimensional Rock--Paper--Scissors model are closely related to the periodic TWs.
Away from the core, the solutions appear to converge asymptotically to a train of periodic TWs. 
These periodic solutions typically have relatively short wavelengths. 
Our results demonstrate that the stability of the two-dimensional SWs can be predicted by the stability analysis of the associated one-dimensional periodic TWs.
This is expected since the onset of instability of SWs coincides with that of the associated periodic TWs \cite{sandstede2000absolute}.

Numerical continuation of spiral waves in reaction-diffusion models has been conducted in \cite{bar2003, barkley1992, Bordyugov2007, Dodson}.
Barkley \cite{barkley1992} developed a numerical method for continuing rotating spiral waves by discretization on an equidistant polar grid.
B{\"a}r et al. \cite{bar2003} performed a pseudo-arclength continuation method to study the existence and stability of spiral waves in a modified Barkley model.
Bordyugov and Engel \cite{Bordyugov2007} described in detail a numerical method for computing and continuing rigidly rotating spiral waves by solving for a large boundary value problem in Fourier space.
Dodson and Sandstede \cite{Dodson} used a continuation scheme to analyze the spectral properties of spiral waves and investigate the underlying mechanisms for instabilities.
Performing continuation-based techniques to study the existence and stability of spiral wave solutions in system~\cref{eq:MeanField} is ongoing work \cite{hasan2020numerical}.

\section*{Acknowledgements} 
The authors thank Stephanie Dodson, Andrus
Giraldo, Edgar Knobloch, Christopher Marcotte, Jens Rademacher, Bj{\"o}rn Sandstede, and Arnd Scheel for fruitful discussions. This project was funded by the Faculty of Science Research Development Fund (FRDF) at the University of Auckland.
CMP is grateful for support from the Marsden Fund Council from New Zealand Government funding, managed by Royal Society Te Ap\={a}rangi, and from the London Mathematical Laboratory.
AMR is
grateful for support from the Leverhulme Trust (RF-2018-449/9).




\begin{thebibliography}{10}


\bibitem{bar2003} B{\"a}r, M., Bangia, A.K. and Kevrekidis, I.G., 2003. Bifurcation and stability analysis of rotating chemical spirals in circular domains: Boundary-induced meandering and stabilization. \textit{Phys. Rev. E}, \textbf{67}, 056126

\bibitem{barkley1992} Barkley, D., 1992. Linear stability analysis of rotating spiral waves in excitable media. \textit{Physical Review Letters}, \textbf{68}, 2090--2093


\bibitem{Bordyugov2007} Bordyugov, G. and Engel, H., 2007. Continuation of spiral waves. \textit{Physica D},  \textbf{228}, 49--58


\bibitem{Busse67} Busse, F.H., 1967. On the stability of two-dimensional convection in a layer heated from below. \textit{J. Math. Phys.},  \textbf{46}, 140--150

\bibitem{Cox2002} Cox, S.M. and Matthews, P.C., 2002. Exponential time differencing for stiff systems. \textit{J. Comput. Phys.},  \textbf{176}, 430--455


\bibitem{Dodson} Dodson, S. and Sandstede, B., 2019. Determining the source of period-doubling instabilities in spiral waves. \textit{SIAM J. Appl. Dyn. Syst.}, \textbf{18}, 2202--2226

\bibitem{Doedel} 
Doedel, E.J., Oldeman, B.E., 2007.
\newblock Auto-07{P}: Continuation and bifurcation software for ordinary differential equations. 
\newblock With major contributions from Champneys, A.R., Dercole, F., Fairgrieve, T.F., Kuznetsov, Yu.A., Paffenroth, R.C., Sandstede, B., Wang, X.J., and Zhang, C.H.; available at \url{http://cmvl.cs.concordia.ca/auto/}


\bibitem{Eckhaus} Eckhaus, W., 1965. \textit{Studies in non-linear stability theory} (Vol. 6). (Springer Science \& Business Media)
  
\bibitem{Frey2010}  Frey, E., 2010. Evolutionary game theory: Theoretical concepts and applications to microbial communities. \textit{Physica A}, \textbf{389}, 4265--4298


\bibitem{Gani2016}  Gani, M.O. and Ogawa, T., 2016. Stability of periodic traveling waves in the Aliev--Panfilov reaction--diffusion system. \textit{Commun. Nonlin. Sci.}, \textbf{33}, 30--42

\bibitem{hasan2020numerical}  Hasan, C.R. and Osinga, H.M. and Postlethwaite, C.M. and Rucklidge, A.M., 2020. Numerical continuation of spiral waves in heteroclinic networks of cyclic dominance. \emph{IMA Appl. Math.} (in press)



\bibitem{Jackson1975}  Jackson, J.B.C. and Buss, L.E.O., 1975. Alleopathy and spatial competition among coral reef invertebrates. \textit{Proceedings of the National Academy of Sciences of the United States of America}, \textbf{72}, 5160--5163

\bibitem{Kerr2002} Kerr, B., Riley, M.A., Feldman, M.W. and Bohannan, B.J., 2002. Local dispersal promotes biodiversity in a real-life game of rock--paper--scissors. \textit{Nature}, \textbf{418}, 171--174

\bibitem{Kirkup2004} Kirkup, B.C. and Riley, M.A., 2004. Antibiotic-mediated antagonism leads to a bacterial game of rock--paper--scissors in vivo. \textit{Nature}, \textbf{428}, 412--414

\bibitem{kr-Lin2008}
Krauskopf, B. and Rie{\ss}, T., 2008
\newblock A {L}in's method approach to finding and continuing heteroclinic connections involving periodic orbits.
\newblock \textit{Nonlinearity}, \textbf{21}, 1655--1690

\bibitem{May1975} May, R.M. and Leonard, W.J., 1975. Nonlinear aspects of competition between three species. \textit{SIAM J. Appl. Math.}, \textbf{29}, 243--253

\bibitem{Nii2000} Nii, S., 2000. The accumulation of eigenvalues in a stability problem. \textit{Physica D}, \textbf{142}, 70--86



\bibitem{om-siads2010}
Osinga, H.M. and Moehlis, J., 2010.
\newblock Continuation-based computation of global isochrons.
\newblock \textit{SIAM J. Appl. Dyn. Sys.}, \textbf{9}, 1201--1228

\bibitem{postlethwaite2017} Postlethwaite, C.M. and Rucklidge, A.M., 2017. Spirals and heteroclinic cycles in a spatially extended rock--paper--scissors model of cyclic dominance. \textit{EPL}, \textbf{117}, 48006

\bibitem{postlethwaite2019} Postlethwaite, C.M. and Rucklidge, A.M., 2019. A trio of heteroclinic bifurcations arising from a model of spatially-extended rock--paper--scissors. \textit{Nonlinearity}, \textbf{32}, 1375--1407

\bibitem{rademacher2007computing} Rademacher, J.D., Sandstede, B. and Scheel, A., 2007. Computing absolute and essential spectra using continuation. \textit{Physica D}, \textbf{229}, 166--183



\bibitem{reichenbach2007} Reichenbach, T., Mobilia, M. and Frey, E., 2007. Mobility promotes and jeopardizes biodiversity in rock--paper--scissors games. \textit{Nature}, \textbf{448}, 1046--1049



\bibitem{romeo2000} Romeo, M.M. and Jones, C.K.R.T., 2000. Stability of neuronal pulses composed of concatenated unstable kinks. \textit{Phys. Rev. E}, \textbf{63}, 011904



\bibitem{sandstede2002stability} Sandstede, B., 2002. Stability of travelling waves. In \textit{Handbook of Dynamical Systems} (Vol. 2) ed B Fiedler (Amsterdam: Elsevier) pp 983--1055


\bibitem{sandstede2000absolute} Sandstede, B. and Scheel, A., 2000. Absolute versus convective instability of spiral waves. \textit{Phys. Rev. E}, \textbf{62}, 7708--7714

\bibitem{sandstede2000Gluing} Sandstede, B. and Scheel, A., 2000. Gluing unstable fronts and backs together can produce stable pulses. \textit{Nonlinearity}, \textbf{13}, 1465--1482


\bibitem{sherratt2013numerical} Sherratt, J.A., 2013. Numerical continuation of boundaries in parameter space between stable and unstable periodic travelling wave (wavetrain) solutions of partial differential equations. \textit{Adv. Comput. Math.}, \textbf{39}, 175--192


\bibitem{Siero2015} Siero, E., Doelman, A., Eppinga, M.B., Rademacher, J.D., Rietkerk, M. and Siteur, K., 2015. Striped pattern selection by advective reaction-diffusion systems: Resilience of banded vegetation on slopes. \textit{Chaos}, \textbf{25}, 036411

\bibitem{Sinervo1996} Sinervo, B. and Lively, C.M., 1996. The rock--paper--scissors game and the evolution of alternative male strategies. \textit{Nature}, \textbf{380}, 240

\bibitem{Sinervo2000} Sinervo, B., Miles, D.B., Frankino, W.A., Klukowski, M. and DeNardo, D.F., 2000. Testosterone, endurance, and Darwinian fitness: natural and sexual selection on the physiological bases of alternative male behaviors in side-blotched lizards. \textit{Horm. Behav.}, \textbf{38}, 222--233

\bibitem{Siteur2014} Siteur, K., Siero, E., Eppinga, M.B., Rademacher, J.D., Doelman, A. and Rietkerk, M., 2014. Beyond Turing: The response of patterned ecosystems to environmental change. \textit{Ecol. Complex.}, \textbf{20}, 81--96

\bibitem{Szczesny2013} Szczesny, B., Mobilia, M. and Rucklidge, A.M., 2013. When does cyclic dominance lead to stable spiral waves? \textit{EPL}, \textbf{102}, 28012

\bibitem{Szczesny2014} Szczesny, B., Mobilia, M. and Rucklidge, A.M., 2014. Characterization of spiraling patterns in spatial rock--paper--scissors games. \textit{Phys. Rev. E}, \textbf{90}, 032704

\bibitem{Szolnoki2014} Szolnoki, A., Mobilia, M., Jiang, L.L., Szczesny, B., Rucklidge, A.M. and Perc, M., 2014. Cyclic dominance in evolutionary games: A review.  \textit{J. Royal Soc. Interface}, \textbf{11}, 20140735


\end{thebibliography}
\end{document}